This is a **postprint** version of:
Díaz-Faes, A.A., Bordons, M. (2017). Making visible the invisible through the analysis of acknowledgements in the humanities. *Aslib Journal of Information Management* 69 (5), 576-590.
The final publication is available at Emerald via https://doi.org/10.1108/AJIM-01-2017-0008

# Making visible the invisible through the analysis of acknowledgements in the humanities

Adrián A. Díaz-Faes*[1], María Bordons[2]

[1]adrian.arias@cchs.csic.es
INGENIO (CSIC-UPV), Universitat Politècnica de València, 46022 Valencia (Spain)
*Corresponding author: adrian.arias@cchs.csic.es. Tel. +34 916022895
[2]maria.bordons@cchs.csic.es
IFS, Spanish National Research Council (CSIC), Albasanz 26-28, 28037 Madrid (Spain)

**Abstract**

**Purpose** — Science is subject to a normative structure that includes how the contributions and interactions between scientists are rewarded. Authorship and citations have been the key elements within the reward system of science, whereas acknowledgements, despite being a well-established element in scholarly communication, have not received the same attention. This paper aims to put forward the bearing of acknowledgements in the humanities to bring to the foreground contributions and interactions that, otherwise, would remain invisible through traditional indicators of research performance.

**Design/methodology/approach** — The study provides a comprehensive framework to understanding acknowledgements as part of the reward system with a special focus on its value in the humanities as a reflection of intellectual indebtedness. The distinctive features of research in the humanities are outlined and the role of acknowledgements as a source of contributorship information is reviewed to support these assumptions.

**Findings** — "Peer interactive communication" is the prevailing support thanked in the acknowledgements of humanities, so the notion of acknowledgements as "super-citations" can make special sense in this area. Since single-authored papers still predominate as publishing pattern in this domain, the study of acknowledgements might help to understand social interactions and intellectual influences that lie behind a piece of research and are not visible through authorship.

**Originality/value** — Previous works have proposed and explored the prevailing acknowledgement types by domain. This paper focuses on the humanities to show the role of acknowledgements within the reward system and highlight publication patterns and inherent research features which make acknowledgements particularly interesting in the area as reflection of the socio-cognitive structure of research.

**Keywords:** Acknowledgements, Reward system, Super-citations, Contributorship, Reward Triangle, Authorship, Humanities, Sociology of Science.

**Paper type**: Research paper

This is a **postprint** version of:
Díaz-Faes, A.A., Bordons, M. (2017). Making visible the invisible through the analysis of acknowledgements in the humanities. *Aslib Journal of Information Management* 69 (5), 576-590.
The final publication is available at Emerald via https://doi.org/10.1108/AJIM-01-2017-0008

## Introduction

Science can be described as a compound of interconnected elements which include: methods through which knowledge can be verified, a cumulative knowledge that emerges from the application of these methods and a set of cultural values and norms that govern scientific activity (Merton, 1973). These norms and values, which have been described as making up the "ethos" of science, are internalised by scientists worldwide. Although the organisational dynamics of research activity may be influenced in each country by local factors, science shows a supra-national character and the norms that govern scientific activity are widely accepted all over the world. Notwithstanding, the behaviour of scientists is not only governed by the normative structure of science, but also by its social structure and the institutionally distinctive reward system. As stated by Merton, the reward system of science was designed to give recognition "to those who have best fulfilled their roles, to those who have best made genuinely original contributions to the common stock of knowledge" (Merton, 1957, p. 639). Accordingly, public contributions to knowledge are rewarded in the form of citations, honorific awards, prizes and academic promotion, all of them conferring prestige and reputation to scientists. In fact, reputation can be assumed as the driving force for building an academic career. In this sense, it can be equalled to the notion of "symbolic capital" introduced by Bourdieu (1988), in such a way that the ranking or status in the academic realm is determined by the possession of prestige which leads scientists to choose the more convenient strategy for its accumulation. Reputation increases the likelihood of scientists to get promoted and to have access to funding and resources, which enable further work. Anyway, the dynamics of knowledge creation involves a constant flow of interactions between scientists that are subject to a number of norms, customs and social factors, which may have an influence on how their contributions are rewarded. As an example, the Matthew effect can be mentioned, according to which well-known scientists tend to get more credit and prestige than less known scientists for the same scientific work (Merton, 1968).

As new knowledge only acquires value when published, scientific publications are the cornerstone of science (Kostoff, 1995). In fact, bibliometric methods shed light on the dynamics of scientific activity and enable to follow the tracks of credit distribution among authors in the publications (authorship) as well as of the recognition they received (citations). Although not free of limitations, both authorship and citations have long proved useful as quantitative methods for analysing different aspects of the research process. Citations provide a measure of the underlying intellectual influence and foundations of research output. It is well-known that a number of factors may affect the probability of being cited such as the exponential growth of research output (more recent publications are more likely to be cited), differences in citation practices by domain, and the availability and the language of publications (see Bornmann and Daniels, 2008). Nevertheless, citations are the best established indicator to estimate the relative importance of a scientific work and have been largely used for the assessment of different research units such as countries, institutions and individuals (Moed and Halevi, 2015). Authorship is a key indicator in the academic reward system (Cronin, 2001), but the allocation of credit to authors is increasingly difficult. Although professional associations and journals have developed criteria and guidelines for authorship, there are no universal criteria for conferring authorship status (Claxton, 2005) or for

This is a **postprint** version of:
Díaz-Faes, A.A., Bordons, M. (2017). Making visible the invisible through the analysis of acknowledgements in the humanities. *Aslib Journal of Information Management* 69 (5), 576-590.
The final publication is available at Emerald via https://doi.org/10.1108/AJIM-01-2017-0008

determining author order in the by-line. As a result, different types of authors' misconduct have been described: individuals who make substantial contributions can be excluded from the author list ("ghost authorship"); while others, who do not contribute significantly, may be listed ("honorary authorship"). Despite the efforts made by journal editors (Wager, 2009), who try to specify formal standards, some evidences reflect the fact that scientists are still barely familiar with authorship criteria (Marusic et al., 2011). On the other hand, the increasing rate of multiple-authorship, which is a consequence of the rising collaborative nature of research, makes even more difficult to identify the specific contribution of a given author to a piece of research. Triggered by the professionalization and institutionalisation of science, which led to the redefinition of how science was done, collaboration burst into science and became a distinctive feature of modern science (Beaver and Rosen, 1979). At present, the increasing specialisation of scientists, together with the need to deal with interdisciplinary research themes and with increasingly expensive facilities have turned research into a highly collaborative activity (Bordons and Gómez, 2000). The number of authors has increased to such an extent that the term "hyper-authorship" is used to denote massive co-authorship levels, mainly associated to the emergence of big science. These practices may undermine the nature and integrity of authorship and create "ethical and procedural difficulties for those responsible for evaluating the nature of individual contributions" (Cronin, 2001, p. 561). The attribution of credit based on contributorship, which suggests including an explicit description of all individual contributions, was proposed in the biomedical journals in the late 1990s to cope with these problems (Rennie *et al.*, 2000). In a contributorship framework, all the contributions to reported work can be identified within an agreed taxonomy of tasks. Although the contributorship model might help to improve the transparency and accuracy of how credit and accountability for scientific work are assigned and recognised, it has been implemented in a limited number of journals (Wager, 2009).

In parallel with the growth in the number of authors, an upward trend in the number of individuals being acknowledged in scholarly journals has also been observed (Cronin, 2001). As it happens with citations and authorship, acknowledgements constitute an important part of formal scholarly communication, but it has not received as much attention as those aforementioned elements (Cronin, 1995). On the one hand, the fact that acknowledgements do not always appear in the same part of the documents (e.g. it may appear before the references or within footnotes in journal articles; as a foreword or preface at the beginning of books or PhD dissertations) represents a difficulty for the collection of data (Giles and Councill, 2004). On the other hand, acknowledgements only recently have started to be included in bibliographic databases, making large-scale analyses possible. Concerning the latter, it should be noted that the acknowledgement section of papers is covered by Web of Science (WoS) since 2008, but only if it is written in English and refers to funding issues. Moreover, acknowledgements are differently covered depending on the database: a) they are collected since 2008 for articles and reviews in the *Science Citation Index Expanded*, b) they have started to be considered since 2015 for all document types in the *Social Sciences Citation Index*, c) whereas *Arts & Humanities Citation Index* has no coverage at all by the time being (see Paul-Hus *et al.*, 2016a for a detailed description). Regarding the Scopus database, funding

This is a **postprint** version of:
Díaz-Faes, A.A., Bordons, M. (2017). Making visible the invisible through the analysis of acknowledgements in the humanities. *Aslib Journal of Information Management* 69 (5), 576-590.
The final publication is available at Emerald via https://doi.org/10.1108/AJIM-01-2017-0008

information is included since 2013, but only funding data is collected and no access to the full text of the acknowledgement section is provided.

Anyway, since WoS database started to collect funding acknowledgements, there has been a renewed interest in acknowledgement studies that is leading to a new branch of literature (see Desrochers *et al*., 2016 for a review). Boosted by the mandate of including funding acknowledgement statements by many funding bodies, most recent efforts to use acknowledgements as indicator of recognition are being focused on their potentialities to trace the output stemming from funding (see for instance, Rigby 2011; Xu *et al*., 2015). But also a few papers have recently explored the existence of specific acknowledgement patterns by domain disclosing and proving its potentiality as source of information on collaboration issues (Costas and van Leeuwen, 2012; Díaz-Faes and Bordons, 2014; Paul-Hus, *et al*., 2016b).

The increasing availability of acknowledgement data has opened new avenues for the study of their function in scholarly communication. Consequently, it also seems the suitable time to reconsider the role that might be played by acknowledgements within the reward system of science. In particular, we believe that acknowledgements can hold a noteworthy symbolic value in the humanities for bringing to the foreground communications and interactions among scientists that model knowledge creation that, otherwise, would remain invisible through traditional indicators. This article is organised as follows: it begins by establishing the value of acknowledgements in the reward system of science as well as their relation with citations and authorship; then the nature and distinctive features of research in the humanities are outlined; and, finally, a literature review is provided to support the special interest of acknowledgements in the area.

## Acknowledgements and the "reward triangle"

Following an implicit code of professional and normative behaviour, acknowledgements appear in scientific literature as voluntary acts of recognition to people whose assistance in the work merits special attention although does not qualify for authorship (Kassirer and Angell, 1991). Acknowledgements are expressions of gratitude regarding financial, editorial, technical, conceptual/cognitive, moral and/or personal support (Cronin, 1995). Among the latter, conceptual/cognitive contributions are particularly relevant because they reflect intellectual indebtedness, which may have a significant bearing on the development of the research. In this sense, it can be mentioned that among the criteria that qualify for authorship, most journal guidelines prioritise intellectual and conceptual contributions over instrumental and technical ones (Wager, 2009; Vinkler, 2010). Acknowledgements may refer to institutions, research councils or government agencies; but also to co-workers, peers or mentors; and even to family members (Kassirer and Angel, 1991; Cronin, 1995). In other words, they "offer insights into the persona of the writer, the patterns of engagement that define collaboration and interdependence among scholars, and the practices of expectation and etiquette that are involved" (Hyland, 2003, p. 244). Thus, naming in the acknowledgement section is no worthless thing; on the contrary, it is a reflection of the cooperative context in which research takes place.

Early literature on acknowledgements, which dates from several decades ago, focuses on the different kinds of tasks that may be recognised in the acknowledgement section, and

This is a **postprint** version of:
Díaz-Faes, A.A., Bordons, M. (2017). Making visible the invisible through the analysis of acknowledgements in the humanities. *Aslib Journal of Information Management* 69 (5), 576-590.
The final publication is available at Emerald via https://doi.org/10.1108/AJIM-01-2017-0008

explores its potentiality as a source of sub-authorship collaboration (Patel, 1973; Heffner, 1981). Nevertheless, it was not until a series of works published by Blaise Cronin and colleagues during the 1990s that the real value of acknowledgements within the reward system was brought to the foreground. In fact, acknowledgements, along with citations and authorship, comprise what has been coined by Cronin and Weaver (1995) as the "reward triangle". As stated critically by these authors, given their social, intellectual and instrumental meaning, as well as their conceptual links with the other two vertices of the triangle, acknowledgements need to be borne in mind in order to obtain a more comprehensive picture of the complex landscape of interactions, influences and collaborations that entails scientific activity.

Figure 1 outlines the position of acknowledgements in the "reward triangle", including their specific features and some that are shared with the other two vertices of the triangle: authorship and citations. Despite being a well-established element of the scholarly communication, acknowledgements clearly differ from authorship and citations in their unstandardised, subtler and more personal nature (Giles and Councill, 2004; Desrochers *et al.*, 2016). As a result, acknowledgements comprise unstructured information which is difficult to analyse. This may explain the fact that they have been under-recognised within the reward system of science. In fact, as stated by Cronin, in the present reward system "the most trivial citation counts for more than the most sincere acknowledgement", which does not seem reasonable (Cronin and Weaver 1995, p. 173). Accordingly, the need to implement measures to make the analysis of acknowledgements possible in a systematic way was pointed out by these authors. However, it should be noted that it is precisely the blurred nature of acknowledgements what offers a unique window into both the traces left by scholarly and non-scholarly communication, and the common practices and cultural norms that characterise each scientific domain. Even though authorship and acknowledgements rely on the work practices that are typical of each domain (Cronin *et al.*, 2003), the latter provides a much more nuance grasp of the interactions that lie behind a piece of research than the mere relation of names appearing in the byline (Díaz-Faes and Bordons, 2014). Following Bourdieu's paradigm of the *homo academicus* (1984), acknowledgements can be deemed as a reflection of the foundations of a discipline and its "ethos".

This is a **postprint** version of:
Díaz-Faes, A.A., Bordons, M. (2017). Making visible the invisible through the analysis of acknowledgements in the humanities. *Aslib Journal of Information Management* 69 (5), 576-590.
The final publication is available at Emerald via https://doi.org/10.1108/AJIM-01-2017-0008

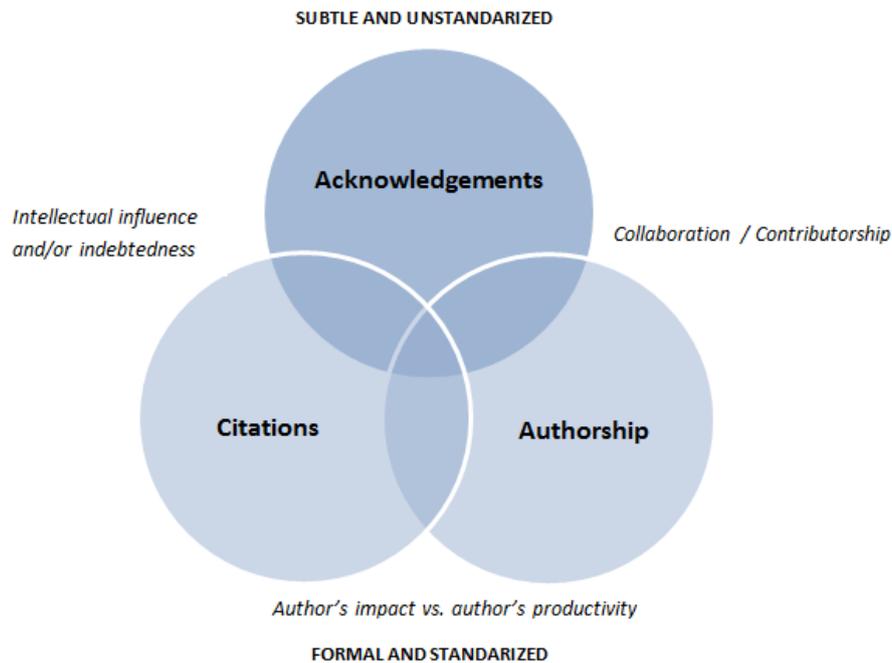

Figure 1. Acknowledgements within the "reward triangle": scope and similarities with citations and authorship.

With regard to the relationship among the different elements of the "reward triangle" a number of considerations can be put forward. Authorship provides a measure of author´s productivity and brings legitimization to scientists, while citations reflect author´s impact on the academic realm. The need to disseminate research through publications is common to all scientists, but not all scientists will be able to produce highly influential and cited research. The interaction between authorship and citations determines scientist's prestige, which is essential for building an academic career. Adding acknowledgements to the reward system of science would contribute to illuminate the other two vertices of the "reward triangle". On the one hand, acknowledgements recognise special assistance of "contributors", who have collaborated in the research but do not meet the requirements to appear as authors — "contributorship side" of the "reward triangle"—. On the other hand, if acknowledgements recognise trusted assessorship (Chubin, 1975), they represent intellectual influence and can be compared to citations —"intellectual influence side"— (Figure 1). Besides, assuming that the contribution of acknowledged individuals might be of greater value than the contribution of some referenced authors, acknowledgements were perceived as "super-citations" by Edge (1979). It should be noted that citations are largely institutionalised as mechanisms to allocate credit among peers since they refer to a document where the influential debt can be trailed; whereas acknowledgements describe private interaction where no solid status can be established (Cronin, 1995). Nevertheless, citations and acknowledgements —particularly those which refer to some kind of intellectual debt— share important features, since both elements appear following a professional code of conduct and leave traces that reflect intellectual and conceptual indebtedness (Cronin and Weaver, 1995). Although no strong correlation has been observed between the number of citations received by researchers and the number of times

This is a **postprint** version of:
Díaz-Faes, A.A., Bordons, M. (2017). Making visible the invisible through the analysis of acknowledgements in the humanities. *Aslib Journal of Information Management* 69 (5), 576-590.
The final publication is available at Emerald via https://doi.org/10.1108/AJIM-01-2017-0008

they are acknowledged, it might indicate that both indicators complement each other and that highly acknowledged researchers are not receiving the deserved recognition (Giles and Councill, 2004). Accordingly, further studies are needed to increase our understanding of the reasons underlying authors' acknowledgements.

### Outlining research in the humanities

To understand the particular interest of the acknowledgement section in the humanities, some inherent features of the area need to be put forward. As far as the research object, the methodology applied and the structure of scholarly communication is concerned, humanities bear some primary differences compared to other areas of research. While an empirical and objective methodology characterises social, natural and life sciences, an interpretative methodology, which focuses on reflective thinking, is applied in the humanities (Gibbons *et al.*, 1994). Both humanities and social sciences seek to study human condition, but they rely on different methodologies to do so. Although humanities research is a heterogeneous domain on account of the manifold set of disciplines and publication patterns that comprise it (Den Hertog, *et al*., 2014), a number of fundamental common characteristics can be outlined (Hemlin and Gustafsson, 1996; Oschner *et al*., 2013). First, research in the humanities is basically focused on theory, source and text. In domains such as philosophy, literature, linguistics or history research tends to be text-driven, so that it is particularly oriented to use original sources which may range from written records in various formats to musical scores, pictures and audio-visual materials. Second, new approaches and criticism are defining features of humanities research in a way that efforts are more focused on providing new reflections and explanations rather than finding new discoveries. Therefore, technical know-how and research facilities are usually not as essential as in other areas, although the use of technology varies considerably by discipline (Borgman, 2007). Special mention deserve some disciplines such as archaeology which is closer to hard sciences in both its discovery oriented nature and the fact that technical know-how is required. Anyway, as a result of the above mentioned characteristics, research in the humanities shows a highly individualistic nature, partly due to the fact that generally it does not require so expensive equipment and lab facilities as research in the natural and experimental disciplines (Gibbons *et al.,* 1994). Finally, due to the fact that humanities knowledge has a deeply rooted influence in society and culture, and plays a significant role for building regional identities, its scope is not restricted to the academia but it reaches the whole society (Whitley, 1984). This means that societal impact is as noteworthy as scientific one.

The above-mentioned features of humanities research have an influence on the communication behaviour of scientists in the area. On the one hand, humanities researchers yield a wide range of different outputs, which comprise many different types of academic – dissemination to peers– and non-academic publications –dissemination to society–. On the other hand, the high rate of individual work that still characterises research in the humanities can be observed through the analysis of its authorship patterns. As an example, Figure 2 shows the authorship pattern for papers (articles and reviews) published by Spanish-based scientists over the past six years according to the WoS database. Around 75% of the publications are single-authored in the humanities, vs. 16% in the social sciences and less than 10% in the

This is a **postprint** version of:
Díaz-Faes, A.A., Bordons, M. (2017). Making visible the invisible through the analysis of acknowledgements in the humanities. *Aslib Journal of Information Management* 69 (5), 576-590.
The final publication is available at Emerald via https://doi.org/10.1108/AJIM-01-2017-0008

remaining domains. Moreover, the co-authorship rate in the humanities remains below two authors per paper in contrast with an extremely high average number of authors in physics – over 61 authors per paper– and, in a lesser extent, in natural and life sciences domains such as clinical medicine and biomedicine –almost 8 authors per paper–. Social sciences also present a more collaborative pattern than humanities –over 3 authors per paper–. Interestingly, the co-authorship index shows an upward trend over the years in all domains, and this also holds for the humanities (from 1.6 authors per paper in 2010 to 1.8 in 2015).

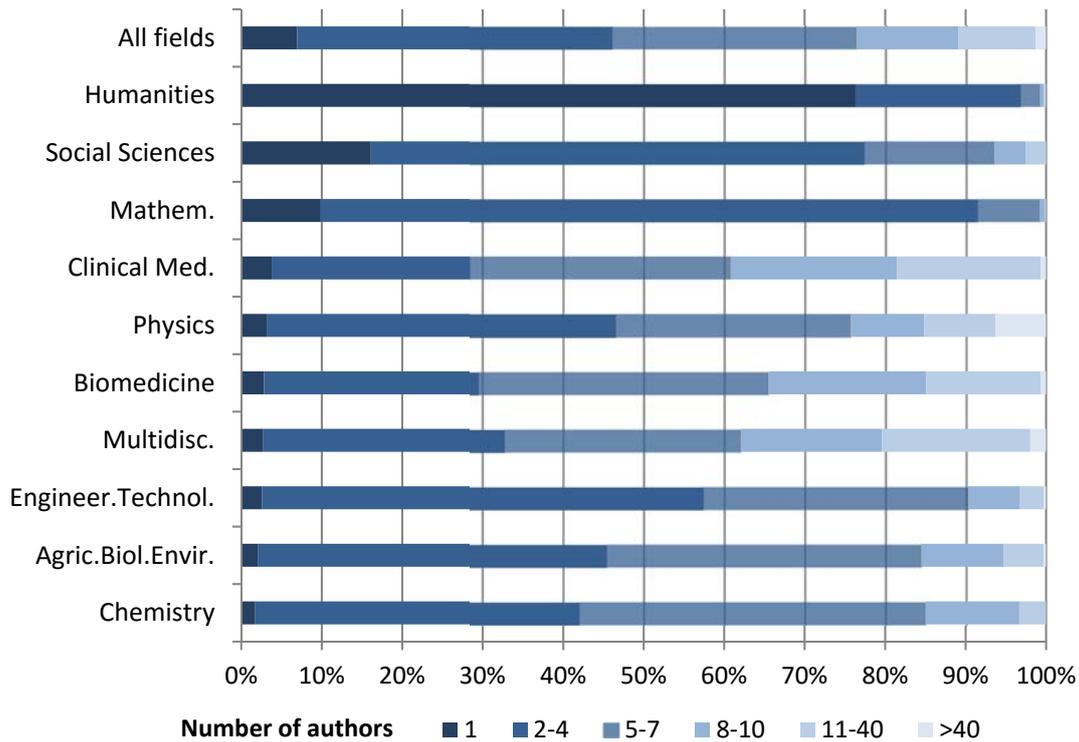

Figure 2. Authorship pattern in the scientific publications of Spain based-researchers (WoS, 2010-2015).
*Note:* only articles and reviews are considered.

The increasing role of collaboration in science has been described at the world level across many scientific fields (Gazni *et al*., 2012), including humanities (Ossenblok *et al*., 2014), where a tendency toward increased teamwork has been observed but with lower growth rates than in the science and engineering fields (Wuchty *et al*., 2007). The fact is that the reach of digital technologies to the humanities has fostered collaborative and interdisciplinary work and is changing scholarship practices (Deegan and McCarty, 2012). However, single-authorship still predominates in the humanities and this has contributed to keep alive the traditional image of researchers from the humanities as "lone wolfs" (Cronin *et al*., 2003). In fact, criticism and debate are integral parts of research in the humanities and writing cannot be separated from the research itself. This also implies that authorship in the humanities is a much more creative and personal act than in the natural and life sciences where writing is more data or reports oriented (Hellqvist, 2010). Nevertheless, even in those theoretically and socially oriented

This is a **postprint** version of:
Díaz-Faes, A.A., Bordons, M. (2017). Making visible the invisible through the analysis of acknowledgements in the humanities. *Aslib Journal of Information Management* 69 (5), 576-590.
The final publication is available at Emerald via https://doi.org/10.1108/AJIM-01-2017-0008

domains where single authored papers still predominate, the idea of a lonely author overlooks the fact that the majority of scientific literature involves communication networks of scholars whether it is formal or informal collaboration (Cronin, 2001; Zuccala, 2006). Interestingly, a recent paper by Paul-Hus *et al.*, (2017) reveals that when acknowledgements are taken into account as a measure of collaboration, the differences traditionally observed between domains in the collaboration rates are much less significant, which implies that some differences may be explained on account of the attribution practices rather than the actual collaboration practices. Therefore, information about sub-authors included in the acknowledgement section may be especially useful in the humanities, where single-authored papers are still the norm and acknowledgements can be the only way to identify social and intellectual underlying interactions.

**Acknowledgements as a source of sub-authorship information**

Since a fundamental feature of humanities research is criticism and discussion, references are used "to evoke a debate or a conversation between lines of thinking" (Hellqvist, 2010, p. 313). A good example of this is the fact that footnotes prevail as a distinctive type of referencing in the humanities due to their function as a source of criticism (footnotes are text comments on the main text). As noted by the above mentioned author, citing might be considered to a certain extent as the main method for the humanities on account of the fact that research is mostly text-driven, but also because there are very few fact making devices compared to natural and life sciences research where figures, graphs or diagrams are used to display empirical findings. In other words, citing can be understood not only as a sign of underlying intellectual influence, but also as a way to build an identity as a researcher where new approaches and theories equate to new empirical findings in the natural and life sciences.

A similar assumption sounds reasonable for the information contained in the acknowledgement section when it comes to intellectual indebtedness. Considering that acknowledgements may reflect socio-cognitive features within a scholarly community, differences in sub-authorship patterns by scientific domain are expected. Table 1 gives an overview of the studies dealing with acknowledgements as a source of collaboration and sub-authorship information with a special focus on the differences by domain[i].

Early studies from 70s and 80s centred their attention on sociological journals and the first division between technical and intellectual support was described (Mackintosh, 1972; Patel, 1973; Heffner, 1981). For instance, Patel (1973) noted tasks such as reading, editing, providing comments to a draft version as kinds of theoretical support. In the 90s, more thorough classifications were proposed (McCain, 1991; Cronin, 1991). The former studied a relatively small set of papers from the biology journal *Genetics* and found that 59% of acknowledgements referred to access to research-related information (technical facilities, experimental materials, unpublished protocols). Among the categories proposed by McCain, "peer interactive communication" (PIC) is worth noting since it gathers a number of tasks that reflect intellectual and conceptual support. This category covers the recognition for providing specific information and critical comments on the manuscript and thanking for advice, discussion and inspiration. In the study of Cronin (1991) six acknowledgement categories were identified: "paymaster" (thank for grants or fellowships), "moral support" (access to facilities),

This is a **postprint** version of:
Díaz-Faes, A.A., Bordons, M. (2017). Making visible the invisible through the analysis of acknowledgements in the humanities. *Aslib Journal of Information Management* 69 (5), 576-590.
The final publication is available at Emerald via https://doi.org/10.1108/AJIM-01-2017-0008

"dogsbody" (support from colleagues in routine work), "technical" (assistance in statistical analysis and computer programming), "prime mover" (provide inspiration or drive) and "trusted assessor" (feedback, comments and critical analysis from peers). Interestingly, the sample was taken from an international journal from library and information science and almost half of the papers thanked "trusted assessors" which can be categorised as PIC support. Subsequent works published by Cronin and colleagues explored the existence of different acknowledgement patterns by domain and showed that PIC acknowledgements were extremely frequent in the humanities and the social sciences domains but with some remarkable differences among disciplines (Cronin *et al.*, 1992; Cronin *et al.*, 1993a; Cronin *et al.*, 1993b; Cronin, 2001; Cronin *et al.*, 2003).

Table 1. Summary of acknowledgement studies focused on contributorship and sub-authorship.

| Article | Sample size, scope | Main acknowledgement patterns |
|---|---|---|
| Mackintosh (1972) | *American Sociological Review,* 1940-65 | A three level typology of acknowledgements is proposed: facilities (i), access to data (ii), and help of individuals (iii) |
| Patel (1973) | N = 7,908 articles from 4 sociological journals, 1895-1965. | Distinguishes between two general types of sub-authorship collaboration: technical assistance and theoretical support |
| Heffner (1981) | N = 395 articles from 28 journals from 4 different domains, 1974-75. Focused on impact of funding | Greater impact of funding upon technical sub-authorship than on theoretical sub-authorship |
| McCain (1991) | N = 241 experimental papers from the journal *Genetics* in 1988 | Typology proposed: access to research-related information (i), access to unpublished, results data (ii), peer interactive communication (iii), technical assistance (iv), manuscript preparation (v). Genetics: 59% of acknowledgements referring to access to research-related information |
| Cronin (1991) | N = 444 articles from the *Journal of the American Society for Information Science,* 1970-90. Focused on PIC acknowledgements | Typology proposed: paymaster (i), moral support (ii), dogsbody (iii), technical (iv), prime mover (v) and trusted assessor (vi). PIC acknowledgements: 47% in library and information science |
| Cronin *et al.*, (1992) | N = 710 articles from 4 top-ranking journals from Library and Information Science, 1979-90. Focused on PIC acknowledgements | PIC acknowledgements: from 46% to 56% in library and information science |
| Cronin *et al.*, (1993a) | N = 307 articles from 4 top-ranking journals from social sciences and humanities, 1971-90 | PIC acknowledgements: 95% in philosophy, 93% in sociology, 84% in history, 78% in psychology |
| Cronin *et al.*, (1993b) | N = 3,081 articles from 10 sociology journals. Focused on PIC acknowledgements | PIC acknowledgements: 73% in sociology |
| Cronin (2001) | N = 909 articles from 5 top journals already analysed in Cronin (1991) and Cronin *et al.*, (1992). Focused on PIC acknowledgements | PIC acknowledgements: 65% in library and information science. More firmly institutionalised than during the 70s and 80s |
| Tiew and Sen (2002) | N = 310 articles from the *Journal of Natural Rubber Research*, 1986-97 | PIC acknowledgements: 44% in polymer science |
| Cronin *et al.*, (2003) | All articles published by the journals *Psychological Review* (n = 1,346) and *Mind* (n = 457), 1900-99 | Philosophy: conceptual (69%). Psychology: financial (36%), conceptual (31%) and instrumental (20%) |
| Cronin *et al.*, (2004) | N = 2,154 articles from the *Journal of the American Chemical Society,* 1900-99 (random sample) | Chemistry: financial (46%), instrumental and technical (34%), conceptual (18%) |
| Hyland (2004) | N = 213 acknowledgement sections from MA (n = 97) and PhD dissertations (n = 113) from six different domains at | Providing access to data was the most common technical support thanked, especially in applied linguistics (75%) and public administration (70%); technical support stands out in |

This is a **postprint** version of:
Díaz-Faes, A.A., Bordons, M. (2017). Making visible the invisible through the analysis of acknowledgements in the humanities. *Aslib Journal of Information Management* 69 (5), 576-590.
The final publication is available at Emerald via https://doi.org/10.1108/AJIM-01-2017-0008

|  | five Hong-Kong universities | biology (38%) and electronic engineering (33%) |
|---|---|---|
| Salager-Meyer *et al.*, (2009) | N = 150 medical papers randomly selected, 2005-07, published by Venezuelan, Spanish and North-American institutions | Clinical medicine: Financial and instrumental/technical supports are the types more acknowledged, being PIC less frequent. Significant differences by country |
| Costas and van Leeuwen (2012) | N = 541,315 world publications, in 2009 in all fields of science<br><br>Focused on PIC acknowledgements<br>PIC identified through a keyword-based search in WoS | PIC acknowledgements: over 50% in economics and, humanities and arts; over 40% in geosciences and multidisciplinary fields; less than 20% in natural and biomedical sciences |
| Díaz-Faes and Bordons (2014) | N=38,257 scientific publications of Spain based-scientists in 2010 in WoS<br>Text patterns in acknowledgements are explored through text-mining in a sample of 1,491 papers from 4 domains | PIC acknowledgements predominate in theoretical or social-oriented fields (statistics and probability, economics), whereas technical assistance is more commonly recognised in experimental research (evolutionary biology). Disclosure of conflict of interest appears in clinical oriented fields (cardiac and cardiovascular systems) |
| Paul-Hus, *et al.*,(2016b) | N = 880,809 world publications in 2014 in all fields of science<br>Text patterns from WoS are explored through text mining techniques | Technical assistance is mostly found in biomedical research, whereas physics, chemistry, and engineering and technology show a more diverse pattern, including specific technical support but also access to facilities and discussions associated with research work<br>PIC acknowledgements related to manuscript improvement are frequently recognised in earth and space and, to a lesser extent, in biology where laboratory work is commonly acknowledged. Potential conflicts of interest emerge in clinical medicine |

*Notes: sample size refers to the number of acknowledgements-bearing papers analysed. PIC = peer interactive communication.*

In the early 2000s, several works pointed out that acknowledgement patterns tend to be more oriented to technical, instrumental and financial support in the natural and life sciences (Tiew and Siew, 2002; Cronin *et al.*, 2004; Hyland, 2004; Salager-Meyer *et al.*, 2009). Since 2008, when WoS began to collect funding acknowledgement data, a number of works have reinforced previous findings by means of much larger sample sizes and fields comparisons showing that PIC acknowledgements are characteristic of theoretical and social oriented domains; whereas natural and life sciences domains are more linked –although not exclusively in some cases–, to technical and laboratory issues as well as to access to facilities and resources (Costas and van Leeuwen, 2012; Díaz-Faes and Bordons, 2014; Paul-Hus *et al.*, 2016b). Since PIC acknowledgements reflect intellectual and conceptual support, they are the ones that can be more precisely equated to the "super-citations" proposed by Edge (Edge, 1979) and their study can be especially appropriate in the humanities, where this type of acknowledgement predominates. In fact, the practice of asking colleagues for comments and criticism on draft papers is more spread in the social sciences and humanities than in the natural and life sciences due to differences in the type of research, but also to the faster publication pace in the latter that prevents authors from asking feedback from colleagues before publication.

## Conclusions

Studying acknowledgements is necessary since they provide valuable information about the interactions, influences and collaborations that take place within the scientific process and, that, otherwise, would be neglected. Hidden aspects of the research process would become

This is a **postprint** version of:
Díaz-Faes, A.A., Bordons, M. (2017). Making visible the invisible through the analysis of acknowledgements in the humanities. *Aslib Journal of Information Management* 69 (5), 576-590.
The final publication is available at Emerald via https://doi.org/10.1108/AJIM-01-2017-0008

visible as well as the contribution of acknowledged individuals, which may deserve recognition. Acknowledgements may recognise technical assistance of individuals who have contributed to the research, but also conceptual support or "peer interactive communication". The latter is the more important type of acknowledgements for identifying intellectual debt to the extent that they have been considered to be at least as valuable as citations (Edge, 1979; Cronin and Weaver, 1995). The notion of acknowledgements as "super-citations", originally prompted by Edge (1979), is certainly a suggestive and rational proposal in the humanities that relies on the assumption that acknowledgements reflect intellectual influence that might be greater than that of some referenced authors. Notwithstanding, there is no empirical evidence of the greater value of PIC acknowledgements as compared to citations within a given piece of research. A potential greater value of PIC acknowledgements could be claimed as a result of their location halfway between authorship and citation. Like citations, they report intellectual influence but, at the same time, there is an interaction between the author and the acknowledged individual who may actively participate in the on-going research. The interactive nature of the influences reported through PIC acknowledgements could make them especially valuable for authors. Further research exploring author's motives for acknowledging and the relevance attributed by authors to the contributions received from acknowledged individuals would shed light on this issue.

Like citations, which reflect influences on the authors work and allow them to "pay" intellectual debts, we consider –as previously suggested by Cronin– that since "peer interactive communication" acknowledgements also reflect intellectual indebtedness, they might be considered within the reward system of science. This can be particularly relevant in the humanities, where PIC is a commonplace due to the nature of the research (text-driven, less importance of technical know-how than in experimental or natural sciences, great significance of originality and criticism) and considering that little information on author interactions is obtained from authorship in this domain where single-authored papers are the norm. Acknowledgements may bring to the foreground important intellectual interactions among scholars such as the mentoring role played by a senior researcher in the course of a PhD, key insight provided by a peer for developing a particular reasoning, critical comments on a manuscript that helped to arise a debate or relevant literature suggestions that allowed a new approach on a subject. Whatever the case, if we assume that PIC acknowledgements may entail significant contributions to the research, it seems reasonable that they get some type of recognition.

Nevertheless, how could acknowledgements receive recognition? The development of an Acknowledgement Index was suggested by Cronin and Weaver (1992), who considered that acknowledgement data could be used conjointly with established bibliometric indicators for research performance assessment. Although the interest and commercial viability of such index was called into doubt by other authors (Baird and Oppenheim, 1994), acknowledgements are now being included in different databases such as WoS or Scopus, opening up a wide range of new possibilities. Even so, the analysis of these data requires the development of special tools to extract acknowledgements from publications, being an example the algorithm elaborated by Giles and Councill (2004) to extract acknowledgements from the CiteSeer digital library. Studies developed up to now have obtained indicators such as

This is a **postprint** version of:
Díaz-Faes, A.A., Bordons, M. (2017). Making visible the invisible through the analysis of acknowledgements in the humanities. *Aslib Journal of Information Management* 69 (5), 576-590.
The final publication is available at Emerald via https://doi.org/10.1108/AJIM-01-2017-0008

the number of acknowledgements per individual, and have revealed that the distribution of acknowledgements to individuals follows a power law —a small number of individuals are acknowledged very frequently, whereas many others are rarely named— (Cronin et al, 1993b; Giles and Councill, 2004). It has been noted that the paucity of acknowledgements earned by the average individual may reduce the usefulness of acknowledgement counting in the context of assessment exercises (Cronin et al. 1993b), even though it does not reduce our interest on them. On the other hand, a number of technical problems such as variant names need to be addressed.

Anyway, using acknowledgement-based indicators with research assessment purposes is not an immediate objective. Prior to that, we need to improve our understanding of acknowledgement practices, which will support decision-making about the interest, convenience and manner of using these indicators in research performance assessments. The need to encourage research on acknowledgements is a basic conclusion that can be drawn from this study. Acknowledgements constitute a long-established element of scholarly communication and their analysis may provide an interesting insight into different aspects of the process of knowledge creation and for understanding social interactions and intellectual relations that lie behind a piece of research. Studying differences between disciplines could be particularly important in the humanities given the fragmented nature of the area as regards their modes of knowledge creation (Whitley, 1984), which might be reflected in the way acknowledgements are used in the different disciplines. Moreover, given the unstandardised and subtle nature of acknowledgements, qualitative research based on the analysis of researcher's views and behaviours would also provide valuable information on acknowledgement patterns in the humanities. As a matter of fact, previous studies have warned of the risks of relying only on computer-based quantitative analysis of acknowledgement texts, since it can be difficult to separate PIC data from other type of recognition —financial, technical, etc.— (Paul-Hus et al., 2016b). Setting measures to foster the standardization of acknowledgement data as far as authors, journals and databases are concerned would contribute to enhance the scope and reliability of the studies.

Citations and authorship have been the cornerstones of the reward system of science, not only due to the valuable information they provide, but also because of their systematic inclusion in international bibliographic databases that began with the creation of the *Science Citation Index* by Garfield in 1964. Since the inclusion of funding acknowledgements in WoS, similar opportunities are open for acknowledgements. Nevertheless, some practical developments should be implemented by both (a) databases and (b) journals to reach these aims. (a) Regarding databases, in particular WoS, some improvements in its acknowledgement indexing policy are needed. Specifically, acknowledgements should be collected in all databases (including the *Arts & Humanities Citation Index*, which has no coverage for the time being); for all documents regardless of their language (only English-written acknowledgements are currently covered) and whether they include or not funding data (only papers with funding data are now covered). In addition, given the fact that books are major communication channels in the humanities, indexing acknowledgements for the *Book Citation Index* would be an important step forward in this sense. (b) With regard to journals, editors should establish separate subsections for different types of information usually jointly recorded under the

This is a **postprint** version of:
Díaz-Faes, A.A., Bordons, M. (2017). Making visible the invisible through the analysis of acknowledgements in the humanities**. *Aslib Journal of Information Management* 69 (5), 576-590.
The final publication is available at Emerald via https://doi.org/10.1108/AJIM-01-2017-0008

acknowledgement section of papers, i.e. financial data, conflict of interest disclosure, and contributorship information (Cronin and Weaver, 1995; Díaz-Faes and Bordons, 2014). This would facilitate the automatic data processing usually done by bibliographic databases and the subsequent analysis of data. Moreover, a widespread implementation of the contributorship model in journals, in addition or instead of the traditional authorship model, might be useful to clarify author´s involvement in research and to identify intellectual contributors. In addition to this, journal guidelines for authors should include clear recommendations concerning acknowledgement policy to ensure that peers are appropriately acknowledged and that only significant contributions are recorded. By taking these measures, shedding new light on the intricate process of knowledge creation in the humanities would be possible and the basis for the study and subsequent recognition of acknowledgements will be provided.

## Acknowledgements

This research was supported by the Spanish Ministry of Economy and Competitiveness (grant CSO2014-57826P).

This is a **postprint** version of:
Díaz-Faes, A.A., Bordons, M. (2017). Making visible the invisible through the analysis of acknowledgements in the humanities. *Aslib Journal of Information Management* 69 (5), 576-590.
The final publication is available at Emerald via https://doi.org/10.1108/AJIM-01-2017-0008

This is a **postprint** version of:
Díaz-Faes, A.A., Bordons, M. (2017). Making visible the invisible through the analysis of acknowledgements in the humanities. *Aslib Journal of Information Management* 69 (5), 576-590.
The final publication is available at Emerald via https://doi.org/10.1108/AJIM-01-2017-0008

This is a **postprint** version of:
Díaz-Faes, A.A., Bordons, M. (2017). Making visible the invisible through the analysis of acknowledgements in the humanities. *Aslib Journal of Information Management* 69 (5), 576-590.
The final publication is available at Emerald via https://doi.org/10.1108/AJIM-01-2017-0008

---

[i] Note that articles included in the table are those which are clearly focused on collaborative patterns beyond funding issues. Due to the difficulties to deal with natural language text appearing in the acknowledgements, works differ strongly in their scope, sample size, units of analysis and the methods through which acknowledgement patterns are explored. For instance, Cronin et al., (1993b) and Costas and van Leeuwen (2012) only focused on PIC acknowledgements, whereas in Cronin et al. (2003) or Díaz-Faes and Bordons (2014) any lexical pattern is explored. Heffner (1981), despite analysing whether funding has greater impact on technical or theoretical sub-authorship

This is a **postprint** version of:
Díaz-Faes, A.A., Bordons, M. (2017). Making visible the invisible through the analysis of acknowledgements in the humanities. *Aslib Journal of Information Management* 69 (5), 576-590.
The final publication is available at Emerald via https://doi.org/10.1108/AJIM-01-2017-0008

collaboration, does not provide real insight about the share of each kind of support. Therefore, comparisons are challenging. These data should be understood as a general overview of the literature on contributorship through acknowledgements analysis.